# Theoretical investigation of Quantum Anomalous Hall Effect in Potassium Tri-vanadium Pentantimonide

*Partha Goswami*
*Physics Dept., D.B. College (University of Delhi), Kalkaji, New Delhi-110019, India*
*Email: physicsgoswami@gmail.com*

**Abstract** The Kagome metal Potassium Tri-vanadium Pent-antimonide can support the quantum anomalous Hall effect theoretically. This is justified by flat bands and Dirac points susceptible to gap opening by spin-orbit coupling or magnetic ordering. The theoretical investigation of this quantum effect is possible exploring strategies like magnetic proximity, and strain or electric gating tuning. Our goal here is to explore the possibility of quantum anomalous Hall effect with a system Hamiltonian involving nearest-neighbour and complex next nearest-neighbour hopping, Rashba spin-orbit coupling, exchange field due to magnetic proximity, and charge density wave. Our preliminary analysis with these ingredients reveals that the system hosts multiple bands whose Chern numbers values suggest weak topological characteristics-not yet quantized, but showing signs of nontrivial Berry curvature accumulation. Upon introducing momentum-space winding, mimicking an orbital magnetic flux, through the momentum-dependence of the phase of the complex hopping, we find that two bands in the multiple band system carry opposite Chern numbers, indicating the emergence of chiral edge states and a quantized anomalous Hall effect. The rest remain trivial, but the system as a whole is no longer topologically inert.

## 1.Introduction

Potassium tri-vanadium pent-antimonide ($KV_3Sb_5$) is an intriguing quantum material that has garnered significant attention due to its rich interplay of topological features, strong electronic correlations, charge ordering phenomena, and chirality **[1–11]**. Structurally, $KV_3Sb_5$ is a layered compound in which vanadium atoms form a kagome lattice—a two-dimensional network of corner-sharing triangles (Figure 1(a))—interspersed with potassium (K) and antimony (Sb) atoms.The crystal structure of $KV_3Sb_5$ adopts the P6/mmm space group, characterized by alternating V–Sb layers separated by K atoms. Within this framework, the vanadium sublattice exhibits a geometrically ideal kagome configuration (Figure 1(b)). As illustrated in Figure 1(c), the antimony atoms are organized into two distinct sublattices: Sb1 forms a simple hexagonal net centered on each kagome hexagon, while Sb2 generates graphene-like sheets positioned above and below the kagome layers. The nearest-neighbor bond lengths are 2.74 Å for both V–V and V–Sb, and 3.16 Å for Sb–Sb. $KV_3Sb_5$ also displays remarkable electronic properties. Its band structure features Dirac cones, van Hove singularities, and flat bands—hallmarks of kagome lattice geometry. Below approximately 78 K, the material undergoes a charge density wave (CDW) transition, and at temperatures below ~1 K, it exhibits superconductivity. At low temperatures, $KV_3Sb_5$ enters a CDW phase which is not a conventional CDW; it breaks inversion symmetry and is believed to be chiral in nature, potentially forming loop current patterns within the charge order.

The charge density wave (CDW) in $KV_3Sb_5$ exhibits spontaneous mirror symmetry breaking, manifesting in left- and right-handed configurations. Its coupling with topological electronic bands opens pathways to unconventional pairing mechanisms, potentially giving rise to chiral superconductivity and the emergence of Majorana edge modes within vortex cores. Superconductivity that breaks time-reversal symmetry (TRS) carries profound implications for quantum computing applications. In $KV_3Sb_5$, the CDW arises from a confluence of electron-phonon interactions, strong electronic correlations, and Fermi surface (FS) nesting. The geometry of the FS facilitates CDW formation through pronounced electron-lattice coupling. Theoretical models **[3, 9-11]** suggest that loop currents can break both inversion symmetry (IS) and TRS, resulting in a chiral CDW state devoid of magnetic moments. The corresponding order parameter is complex, incorporating significant imaginary components. The orbital-selective currents lead to anomalous Hall effect (AHE), even in the absence of net magnetization. There is compelling evidence for broken TRS and/or IS in $KV_3Sb_5$**[5].** Scanning tunneling microscopy reveals bond-order modulations with rotational asymmetry in $KV_3Sb_5$, indicative of bond-centered CDW patterns that violate the sixfold ($C_6$) rotational symmetry **[12]**.

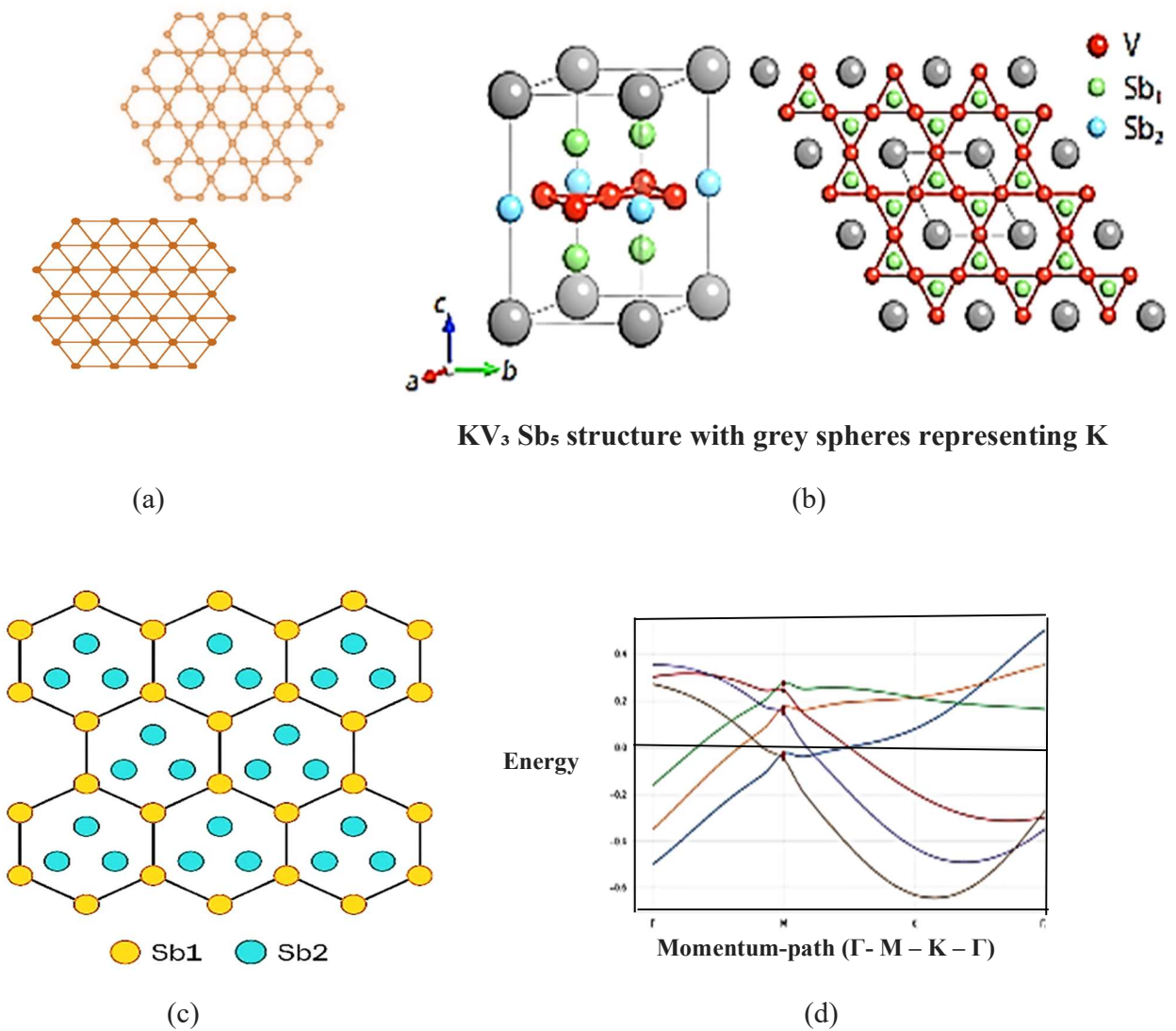

**Fig.1. (a)** A visual representation of non-chiral Kagome lattice and two-dimensional network of corner-sharing triangles. **(b)** The $KV_3Sb_5$ prototype structure ( P6/mmm space group) is shown featuring a layered arrangement of V-Sb sheets separated by K (Grey spheres), with the vanadium sublattice forming a structurally perfect Kagome lattice. **(c)** Two distinct Sb sublattices are present, with the Sb1 sublattice forming a simple hexagonal net centred on each

Kagome hexagon, and the Sb2 sublattice creating a graphene-like Sb sheet above and below each Kagome layer. **(d)** We use a standard kagome unit cell with three sublattices (A, B, C) and expanded to spinful basis (↑, ↓) to form a 6×6 minimal Bloch Hamiltonian below. The nearest neighbor (NN) plus complex next-nearest neighbor (NNN) terms, Rashba spin-orbit coupling (RSOC), and charge density wave (CDW) parameter and were implemented in a compact, phenomenological way. The sketch of the expected six-band structure with these ingredients shows spin-split flat bands near the Fermi level (represented by the horizontal solid line). The saddle points near the M point are highlighted in red—these are regions where the band curvature changes, leading to van Hove singularities and enhanced electronic interactions. The almost linear energy dispersion (Dirac cones) near the M point in the Brillouin zone is visible in the plot.

Can a Kagome metal (like $KV_3Sb_5$) support the quantum anomalous Hall effect (QAHE)? Theoretically, the answer is affirmative **[13]**. The justification for this lies in the following points: (a) The Kagome lattice in $KV_3Sb_5$ features flat bands and Dirac points that are susceptible to gap opening by spin-orbit coupling (SOC) or magnetic ordering, which are crucial elements for QAHE. (b) $KV_3Sb_5$ is known to exhibit strong SOC due to its heavy Sb atoms; if TRS is broken (for example, through intrinsic or induced magnetic order), a QAHE-like phase could potentially arise. (c) Although QAHE has not been observed, the presence of anomalous Hall signals, chiral charge order, and possible time-reversal symmetry breaking below the CDW transition (~80 K) suggests that the necessary conditions for QAHE are met. Its realization has yet to be achieved due to various theoretical and practical obstacles. Specifically, the following challenges exist: (a) $KV_3Sb_5$ does not exhibit intrinsic magnetism in the conventional sense. TRS breaking may occur due to orbital or CDW-induced effects, rather than magnetic spins. (b) QAHE requires the emergence of a bulk gap at Dirac points due to magnetic interactions, which has not been clearly observed in $KV_3Sb_5$. (c) QAHE necessitates clean systems, band isolation, and often fine-tuned chemical potentials. Real-world $KV_3Sb_5$ samples may be affected by excessive disorder or imperfect band alignment. As QAHE remains a goal for Kagome metals, our primary objective in this paper is to investigate this issue theoretically. To this end, we start with a standard kagome unit cell with three sublattices (A, B, C) and expanded to spinful basis (↑, ↓) to form a 6×6 minimal Bloch Hamiltonian. The nearest neighbor (NN) plus complex next-nearest neighbor (NNN) terms, Rashba spin-orbit coupling (RSOC), and charge density wave (CDW) parameter and were implemented in a compact, phenomenological way. The artist's impression of the expected six-band structure derivable from this minimal Hamiltonian is shown in Fig.1d. The figure shows spin-split flat bands near the Fermi level (represented by the horizontal solid line), and somewhat sharpened van Hove singularities near the M and K-points, manifesting as a distinct peak indicative of enhanced density of states due to band curvature. The almost linear energy dispersion (Dirac cones) near these high-symmetry points in the Brillouin zone are visible in the plot. This is due to CDW formation**,** and spin-orbit coupling**.** The Dirac bands are asymmetric meaning splitting of the bands**,** that is the originally degenerate energy levels separate into distinct bands with a measurable energy gap between them. The possible strategies for investigating QAHE include (a) inducing magnetic proximity from a substrate (MPS), (b) doping with magnetic atoms (e.g., Mn, Cr), and (c) tuning via strain or electric gating. We employ the first approach in this problem by introducing an MPS-induced exchange field term. This term breaks TRS mimicking an out-of-plane magnetization. It is essential to recognize that, whereas AH conductance is continuous and non-quantized, QAH conductance is quantized in integer multiples of $\left(\frac{e^2}{h}\right)$. The former is not necessarily topological, whereas the latter is topologically nontrivial with a non-zero Chern number (C). The latter stems from a topologically nontrivial band structure and intrinsic magnetism. Conversely, the former arises from Berry curvature (BC), intrinsic/ extrinsic spin-orbit coupling, and magnetization. Upon employing the Fukui-Hatsugai- Suzuki (FHS) method [14-17], we compute the Chern numbers for each band (from lowest to highest energy). Furthermore, upon introducing momentum-space winding, mimicking an orbital magnetic flux, through the momentum-dependence of the phase of the complex hopping, we find that two bands in the multiple band system carry opposite Chern numbers (one of the two occupied bands produces a Chern number of $C \approx +1$, whereas the other produces $C \approx -1$ and, thus, these bands demonstrate topological non-triviality), indicating the emergence of chiral edge states and a quantized anomalous Hall effect. The rest of the bands exhibit trivial topology, characterized by a Chern number of $C \approx 0$ in each instance, but the system as a whole is no longer topologically inert. This scenario is unlike other Kagome systems such as $Co_3Sn_2S_2$ which is closer to realizing QAHE [18] with overall Chern number quantized. It follows that achieving the quantum anomalous Hall effect (QAHE) in $KV_3Sb_5$ seems attainable either by electron or hole doping, which adjusts the Fermi level to populate bands possessing non-zero Berry curvature. Another promising strategy involves designing superlattices with other two-dimensional materials, such as transition metal dichalcogenides (TMDs), to effectively tailor the band structure. We have discussed this issue in brief in Sect.4..

The paper is organized in the following manner: In Sect. 2, we present the model of $KV_3Sb_5$ with all essential ingredients mentioned above. We, in fact, provide a model of the Bloch Hamiltonian matrix in momentum space for a Kagome lattice of $KV_3Sb_5$ with one orbital, one layer, three sublattices (A, B, C), and two spin states (↑, ↓). It helps visualize the structure without being overly large. In Sect. 3, we obtain the corresponding band structure and calculate QAH conductance. We present future perspective and very brief concluding remarks in Sect. 4.

## 2. Model

To formulate a comprehensive tight-binding Hamiltonian (TBH) for the compound $KV_3Sb_5$, one must incorporate the intricate interplay of multiple physical phenomena and orbital degrees of freedom. This includes contributions from vanadium 3d and antimony 5p orbitals, spin-orbit coupling (SOC), mean-field electron-electron interactions that enable self-consistent charge density wave (CDW) formation, and interlayer hopping processes. The formulation must reflect the quasi-two-dimensional nature of $KV_3Sb_5$ and the influence of its in-plane Kagome lattice geometry. A three-dimensional real-space Hamiltonian is constructed to capture the essential physics of $KV_3Sb_5$. It includes nearest-neighbour (NN) and next-nearest-neighbour (NNN) hopping terms, with the latter incorporating time-reversal symmetry (TRS) breaking effects. The model also integrates SOC and interlayer coupling between stacked Kagome layers along the crystallographic c-axis. These features collectively give rise to two-dimensional electronic instabilities, such as Fermi surface (FS) nesting and van Hove singularities. The presence of the singularities—characterized by sharp peaks in the density of states arising from saddle points in the band structure—enhances electronic correlations in $KV_3Sb_5$. These correlations are believed to play a pivotal role in the emergence of unconventional and potentially topological superconductivity. Such phenomena may ultimately facilitate the realization of Majorana zero modes, offering promising avenues for quantum information processing and topological quantum computing.

The KV$_3$Sb$_5$ unit cell consists of three V atoms (sublattices A, B, and C), each contributing d-orbitals, primarily $d_{xy}$, $d_{xz}$, and $d_{yz}$. These three $d$-orbitals display odd inversion symmetry and transform in a manner analogous to vectors under the point group symmetries of the Kagome lattice. This property makes them conducive to in-plane hopping between V atoms within the Kagome network. Furthermore, the treatment of Sb atoms involves the utilization of effective p-orbitals, which may be either explicitly included or integrated out. Additionally, the consideration of spin effectively doubles the degrees of freedom. It is worth noting that the orbitals $d_{z^2}$, and $d_{x^2-y^2}$ exhibit $e_g$ symmetry, a characteristic typically associated with elevated energy levels in environments resembling octahedral or trigonal prismatic structures. Consequently, the basis vector, in terms of annihilation operators, is $\Psi = (c_{A,\alpha,\sigma}, c_{B,\alpha,\sigma}, c_{C,\alpha,\sigma})^T, \sigma = (\uparrow,\downarrow)$, where each $c_{i,\alpha,\sigma}$ represents a multi-orbital vector at site $i \in (A, B, C)$. Considering nearest-neighbour (NN) hopping $t_{i,j}$ and next-nearest-neighbour (NNN) hopping $t'_{i,j}$ between orbitals ($d_{xy}$, $d_{xz}$, and $d_{yz}$) $\alpha = m, n$ on sites $i$, and $j$, the corresponding term in TBH can be written as $H_{hop} = [\sum_{\langle im,jn \rangle, \sigma=(\uparrow,\downarrow)} t_{im,jn} c^\dagger_{im\sigma} c_{jn\sigma} + \sum_{\langle\langle im,jn \rangle\rangle, \sigma=(\uparrow,\downarrow)} t'_{im,jn} e^{i\phi} c^\dagger_{im\sigma} c_{jn\sigma} + \text{h.c.}]$, where hopping respect $C_6$ rotation and reflection. The NNN hopping includes complex phase $e^{i\phi}$ to mimic loop current. We will first consider the case $\phi$ constant. Next, we will assume $\phi = \phi(\mathbf{k})$ – the momentum dependence mimicking an orbital magnetic flux. The TRS breaking term comes from complex $t'_{i,j}$. Another source of TRS breaking term is MPS discussed above. The corresponding term could be written as $H_{MPS} = [\sum_{i,m,\sigma=(\uparrow,\downarrow)} J c^\dagger_{im\sigma} c_{im\sigma}]$ where $J$ is the strength of the proximity-induced exchange field. It must be noted that for investigating QAHE, a boosted out of plane magnetization is crucial (Sect. 3). The intrinsic Kane-Mele (KM) type SOC terms on V sites can be expressed as $H_{KM} = i\lambda_{SO} \sum_{\langle\langle im,jn \rangle\rangle, \sigma=(\uparrow,\downarrow),\sigma'} \nu_{i,j} c^\dagger_{im\sigma} (s_z) c_{jn\sigma'}$, where $\lambda_{SO}$ denotes the KM-SOC strength (weak to moderate for V and strong for Sb), $\nu_{i,j} = \pm 1$ encodes the chirality of the path from $j$ to $i$, and $(s_z)$ is the Pauli matrix. For each NNN pair $\langle\langle i, j \rangle\rangle$ we compute $\nu_{i,j} = \pm 1$ depending on whether the path turns left or right at the intermediate site. The Kane-Mele SOC (IS is preserved and sublattice localization respected) is often used as a minimal model interaction for opening topological bandgaps at Dirac points like in graphene or Kagome lattice. It must be noted that $H_{KM}$ breaks spin $SU(2)$ but preserves TRS. In real materials like KV$_3$Sb$_5$, inversion symmetry is broken either explicitly through charge-density waves, surface terminations, substrate effects, or built-in fields, or spontaneously via time-reversal or spatial symmetry breaking in electronic or lattice order. So, Rashba SOC(RSOC) becomes important as it originates from inversion symmetry breaking at the interface. It induces spin splitting even at TRS-invariant momenta (e.g., $\Gamma$ point), and leads to non-conserved spin and helical spin textures. The RSOC Hamiltonian, in tight-binding form for Kagome, is $H_{RSOC} = i\lambda_R \sum_{\langle im,jn \rangle, \sigma=(\uparrow,\downarrow)} c^\dagger_{im\sigma} (\mathbf{s} \times \widehat{d_{ij}})_z c_{jn,-\sigma} + \text{h.c.}$, where $\lambda_R$ is the RSOC strength, $\widehat{d_{ij}} = (\mathbf{r}_i - \mathbf{r}_j)/|(\mathbf{r}_i - \mathbf{r}_j)|$ is the unit vector between sites, $\mathbf{s}$ corresponds to Pauli matrices, and $(\mathbf{s} \times \widehat{d_{ij}})_z = (s^x \widehat{d^y_{ij}} - s^y \widehat{d^x_{ij}}) \exp(i\mathbf{k} \cdot \boldsymbol{\delta}_{ij})$. The Hamiltonian $H_{RSOC}$ breaks inversion symmetry. Here, $\mathbf{r}_i$ is the position of the sublattice site $i \in (A, B, C)$, $\boldsymbol{\delta}_{ij}$ is the vector site $j$ to $i$, and $\mathbf{k} = (k_x, k_y)$ is the crystal momentum. The pertinent inquiry is "which (Kane-Mele or RSOC) to apply in KV$_3$Sb$_5$?". The response is, if one is simulating the topological insulating gap at Dirac points in a symmetric Kagome lattice and wishes to employ a simplified, spin-conserving model to explore Berry curvature and Z$_2$ topology, one requires the utilization of $H_{KM}$. Conversely, if one is investigating spin-split bands, helical spin textures, non-centrosymmetric CDW states, surface states, or exploring the possibility of QAHE one requires the utilization of $H_{RSOC}$. As we shall see below, the successful exploration requires cranked up $\lambda_R$ (Sect. 3). In KV$_3$Sb$_5$, possibly this is requisite, due to the strong SOC resulting from heavy V and Sb atoms. Additional reasons encompass broken inversion symmetry (IS) from CDW phases and the investigation of topological features, e.g., Dirac points, and potential quantum spin Hall (QSH) behaviour.

We model the electron-electron interaction as on-site density-density repulsion, treated in mean-field approximation as $H_{CDW} = \sum_{im,\sigma} U \langle \widehat{d_{im}} \rangle c^\dagger_{im\sigma} c_{im\sigma}$, where the CDW ansatz impose modulated density as $\langle \widehat{d_{im}} \rangle = d_0 + \delta \cos(r_i \cdot Q)$. Here the Bravais lattice vectors are $a_1 = a(1,0)$ and $a_2 = a\left(\frac{1}{2}, \frac{\sqrt{3}}{2}\right)$. The position $r_i$ of $(A, B, C)$ relative to the unit cell origin are typically chosen as $r_A = a(0,0), r_B = \frac{1}{2} a_1 = a\left(\frac{1}{2}, 0\right)$, and $r_C = \frac{1}{2} a_2 = a\left(\frac{1}{4}, \frac{\sqrt{3}}{4}\right)$. A single- $Q$ CDW vector is $\left(\frac{2\pi}{3a}\right)(1,0)$, while double- $Q$ CDW vectors are $\left(\frac{2\pi}{3a}\right)(1,0)$, and $\left(\frac{2\pi}{3a}\right)\left(-\frac{1}{2}, \frac{\sqrt{3}}{2}\right)$. The typical triple CDW vectors, commonly observed vectors in KV$_3$ Sb$_5$, are $Q_1 = \left(\frac{2\pi}{3a}\right)(1,0)$, $Q_2 = \left(\frac{2\pi}{3a}\right)\left(-\frac{1}{2}, \frac{\sqrt{3}}{2}\right)$, and $Q_3 = \left(\frac{2\pi}{3a}\right)\left(-\frac{1}{2}, -\frac{\sqrt{3}}{2}\right)$. It forms a chiral representation [12] for KV$_3$Sb$_5$ under the irreducible representations of the $C_{3z}$ point group which is a subgroup of the point group $D_{6h}$. The compound KV$_3$Sb$_5$ [6] has a layered Kagome lattice with the group $D_{6h}$ in its undistorted state. But below the CDW transition temperature ~78 K, when TRS braking is manifested leading to anomalous Hall/Nernst effects, the symmetry is lowered (C$_6$ → C$_2$) due to lattice distortions. The three modulation wave vectors ( $Q_1, Q_2, Q_3$) in KV$_3$Sb$_5$ are 120° apart in reciprocal space and have equal magnitudes. This is exactly what is expected from a symmetric triple-**Q** modulation, which respects the three-fold rotational symmetry of the lattice but breaks mirror symmetry, eventually leading to chirality. How? If all three modulations are in phase, the resulting CDW is non-chiral and preserves TRS. But if the modulations have relative phase shifts — say, one lags or leads the others — the superposition leads to orbital loop currents in real space circulating around the kagome triangles that break TRS**.** The TRS breaking means the system distinguishes between clockwise and counterclockwise processes. The resulting charge pattern is not mirror symmetric, giving rise to chirality. Thus, the phase difference discussed above introduces a handedness to the charge pattern — a chiral CDW**.** The scalar triple product ( $Q_1 \times Q_2) \cdot Q_3$) is non-zero which further emphasizes chirality in **k**-space. Once again, how? The currents circulate in a preferred direction which is encoded by the scalar triple product and the sign of the product tells the handedness: positive means right-handed orientation, negative means left-handed. This orientation cannot be changed by simple rotation—it requires reflection, which is the essence of chirality. Upon considering the three modulation wave vectors $Q_j = (Q_1, Q_2, Q_3)$, the simplified expression for the CDW term is $\sum_{\alpha,m,\sigma} \Delta_{\alpha,CDW}(r_\alpha) c^\dagger_{\alpha m\sigma} c_{\alpha m\sigma}$, where $\alpha \in (A, B, C)$, where the CDW order parameter $\Delta_{\alpha,CDW}(r_\alpha) = \sum_{j=1}^{3} \Delta_\alpha e^{i(Q_j \cdot r_\alpha + \phi_j)}$. The order parameter here comprises both real and imaginary components. The imaginary part induces circulating currents or orbital magnetization, which are odd under time-reversal symmetry (TRS). When considering only the real component—particularly in configurations where the phases $\phi'_j s$ are equal —the resulting CDW pattern is symmetric and achiral. However, when the phases differ by fixed increments (e.g., $\phi_j = 0, 2\pi/3, 4\pi/3$), the superposition of CDWs generates a helical or corkscrew-like modulation in real space. This non-coplanar phase arrangement breaks mirror symmetry, introducing chirality. Such chirality inherently violates TRS and can give rise to effects such as the anomalous Hall effect. The interplay between the real and imaginary components of the charge density wave (CDW) order parameter in KV$_3$Sb$_5$ gives rise to chiral CDW states exhibiting Z$_3$ nematicity, wherein the system spontaneously selects one of three symmetry-equivalent lattice directions. This emergent chirality is not only intrinsic but also magnetically switchable under moderate external fields, owing to the coupling between the magnetic field and the orbital currents associated with the chiral CDW. Such coupling facilitates reversible transitions between clockwise and counterclockwise chiral configurations. The coexistence of chiral CDW order, Z$_3$ nematic symmetry breaking, and field-tunable chirality positions KV$_3$Sb$_5$ as a versatile platform for investigating a rich landscape of quantum phenomena. These include topological phases, unconventional superconductivity, and electronic symmetry breaking, where collective electronic motion can be dynamically reoriented through the application of magnetic flux. This tunability underscores the potential of KV$_3$Sb$_5$ in advancing quantum materials research and may hold implications for future technologies based on topological control and quantum coherence.

The interlayer hopping between stacked Kagome layers along the crystallographic c-axis plays a significant role in the electronic structure, particularly for the $d_{z^2}$ orbital. This vertical coupling, if comparable in magnitude to the dominant in-plane hopping processes, must be incorporated explicitly. The corresponding contribution to the Hamiltonian is expressed as: $H_z = [\sum_{m,n\langle i,j\rangle_z \sigma} t^z_{m,n} c^\dagger_{im\sigma} c_{jn\sigma}]$, where $t^z_{m,n}$ interlayer hopping with $(m,n)$ indexing the relevant orbitals. In this paper we will not consider this term relegating it to a future communication. Therefore, the total Hamiltonian is $(H_{hop} + H_{SOC} + H_{CDW} + H_{MPS})$. In matrix form, it is a $N \times N$ Bloch Hamiltonian, where $N$ = Number of sublattices $\times$ Spin $\times$ Number of orbitals $\times$ Number of layers. This formalism enables the study of band structure, topological features, and interaction-driven phenomena arising from the intricate coupling between layers in Kagome-based systems.

A structured representation of the total Hamiltonian $(H_{ho} + H_{SOC} + H_{CDW} + H_{MPS})$ in momentum space, incorporating nearest-neighbour (NN) and next-nearest-neighbour (NNN) hopping, spin-orbit coupling (SOC), CDW parameter, and interlayer hopping, is given below. We'll focus first on the basis definition followed by the non-interacting part for the Bloch Hamiltonian, suitable for band structure calculations. For the former, we assume the following:(a) Kagome lattice with 3 sublattices: A, B, C.(b) 2 spin states: ↑,↓. (c) three orbitals ($d_{xy}$, $d_{xz}$, and $d_{yz}$).(d) Two layers stacked along the c-axis. The basis set in momentum space is $\Psi_k = (c^{A1}_{k,\uparrow} c^{B1}_{k,\uparrow} c^{C1}_{k,\uparrow} c^{A1}_{k,\downarrow} c^{B1}_{k,\downarrow} c^{C1}_{k,\downarrow} \ldots\ldots c^{C2}_{k,\downarrow})^T$. Each entry accounts for sublattice (A, B, C), layer index (1 or 2), spin orientation, and orbital state (implicitly included per component if needed). The total basis size is, therefore, $N = 36$. We provide below an explicit example of the Bloch Hamiltonian matrix in momentum space for a Kagome lattice of KV$_3$Sb$_5$ with one orbital, one layer, three sublattices (A, B, C), and two spin states (↑, ↓). It helps visualize the structure without being overly large. This setup gives a total basis size of six in $k$-space: $\Psi_k = (c^A_{k,\uparrow} \ c^B_{k,\uparrow} \ c^C_{k,\uparrow} \ c^A_{k,\downarrow} \ c^B_{k,\downarrow} \ c^C_{k,\downarrow})^T$. Let's consider these terms: NN hopping of amplitude $t$, NNN Hopping of amplitude $t'$, SOC of amplitude $\lambda_R$ with imaginary off-diagonal terms (only affects NNN in Kagome systems), and the simplified expression for the CDW term. This is a Hermitian foundational model and may be used to study topological phases and flat-band physics.

$$H(\mathbf{k}) = \begin{bmatrix} \Delta_A + h_A & u_{AB} & u_{CA}^* & 0 & v_{AB} & -v_{CA}^* \\ u_{AB}^* & \Delta_B + h_B & u_{BC} & -v_{AB}^* & 0 & v_{BC} \\ u_{CA} & u_{BC}^* & \Delta_C + h_C & v_{CA} & -v_{BC}^* & 0 \\ 0 & -v_{AB} & v_{CA}^* & \Delta_A + h_A & u_{AB} & u_{CA}^* \\ v_{AB}^* & 0 & -v_{BC} & u_{AB}^* & \Delta_B + h_B & u_{BC} \\ -v_{CA} & v_{BC}^* & 0 & u_{CA} & u_{BC}^* & \Delta_C + h_C \end{bmatrix} \quad (1)$$

These terms introduce band bending and shift Dirac/Weyl crossings depending on the symmetry of the phase. The Hamiltonian expectedly captures all the essential physics of the single-layer Kagome system: including chirality, topological band inversion, and SOC-driven degeneracy lifting. Here, Kagome lattice has 3 sites per unit cell, include hopping between nearest neighbours (A↔B, B↔C, C↔A), SOC only couples spins on NNN paths with complex phases, and CDW introduces on-site energy variation, breaking translational symmetry but treated here in momentum space as diagonal terms. The CDW terms ($\Delta_A, \Delta_B, \Delta_C$), in their simplest form, correspond to on-site energies for the sublattices (A, B, C). The NN hopping terms for model of the metal KV$_3$Sb$_5$ are $u_{AB}(\mathbf{k}) = t(1 + e^{i\mathbf{k}\cdot\boldsymbol{\delta}_{AB}})$, $u_{BC}(\mathbf{k}) = t(1 + e^{i\mathbf{k}\cdot\boldsymbol{\delta}_{BC}})$, and $u_{CA}(\mathbf{k}) = t(1 + e^{i\mathbf{k}\cdot\boldsymbol{\delta}_{CA}})$. The exponential terms capture the hopping to the neighbour connected via vector $\boldsymbol{\delta}_{ij}$, preserving Bloch periodicity. The Kagome lattice consists of corner-sharing triangles. Given the Bravais lattice vectors are $\mathbf{a}_1 = a(1,0)$ and $\mathbf{a}_2 = a\left(\frac{1}{2}, \frac{\sqrt{3}}{2}\right)$, then what are the NN vectors are $\boldsymbol{\delta}_{AB}, \boldsymbol{\delta}_{BC}$, and $\boldsymbol{\delta}_{CA}$? We define the NN vectors so they connect the three sites of the triangle (A→B, B→C, C→A) symmetrically: $\boldsymbol{\delta}_{AB} = \left(\frac{a}{2}\right)(1,\sqrt{3}), \boldsymbol{\delta}_{BC} = \left(\frac{a}{2}\right)(-1,\sqrt{3})$, and $\boldsymbol{\delta}_{CA} = a(0,-1)$. The NNN hopping corresponds to $h_\alpha$'s where $\alpha \in (A,B,C)$. The NNN hopping terms can and often should be complex, especially in Kagome systems exhibiting topological features. In fact, in a typical Kagome lattice, the second-neighbour paths form closed loops (e.g. hexagons), and introducing complex phase factors (typically from orbital flux or effective magnetic fields) breaks time-reversal or inversion symmetry, enabling phenomena like topological band structures (e.g. Chern insulators), band splitting and flat bands, and non-trivial Berry curvature. Consider now sublattice A and vectors $\boldsymbol{\delta}_i$. These vectors point to NNN sites around hexagon centred on A sites. We have $\boldsymbol{\delta}_1 = \mathbf{a}_1 + \mathbf{a}_2 = a\left(\frac{3}{2}, \frac{\sqrt{3}}{2}\right)$, $\boldsymbol{\delta}_2 = -\mathbf{a}_1 + \mathbf{a}_2 = a\left(-\frac{1}{2}, \frac{\sqrt{3}}{2}\right)$, and $\boldsymbol{\delta}_3 = -\mathbf{a}_2 = a\left(-\frac{1}{2}, -\frac{\sqrt{3}}{2}\right)$. For sublattice B, the corresponding vectors $\boldsymbol{\delta}_i'$ are given by $\boldsymbol{\delta}_1' = a(-1,0), \boldsymbol{\delta}_2' = \left(\frac{a}{2}\right)(1,-\sqrt{3})$, and $\boldsymbol{\delta}_3' = \left(\frac{a}{2}\right)(1,\sqrt{3})$. These are 120° rotated version of A vectors. Similarly, for the sublattice C rotated again by another 120°, the corresponding vectors $\boldsymbol{\delta}_i''$ are given by $\boldsymbol{\delta}_1'' = \left(\frac{a}{2}\right)(1,-\sqrt{3}), \boldsymbol{\delta}_2'' = \left(\frac{a}{2}\right)(3,\sqrt{3})$, and $\boldsymbol{\delta}_3'' = a(-1,0)$. These are NNN displacement vectors from a site on a sublattice to one of its NNN partners. Mathematically, this means NNN hopping terms take the form

$$h_A(\mathbf{k}) = t'\left[e^{i\left(\mathbf{k}\cdot\boldsymbol{\delta}_1 + \frac{\pi}{2}\right)} + e^{i\left(\mathbf{k}\cdot\boldsymbol{\delta}_2 - \frac{\pi}{2}\right)} + e^{i\left(\mathbf{k}\cdot\boldsymbol{\delta}_3 + \frac{\pi}{2}\right)}\right] - \mu,$$

$$h_B(\mathbf{k}) = t'\left[e^{i\left(\mathbf{k}\cdot\boldsymbol{\delta}_1' - \frac{\pi}{2}\right)} + e^{i\left(\mathbf{k}\cdot\boldsymbol{\delta}_2' + \frac{\pi}{2}\right)} + e^{i\left(\mathbf{k}\cdot\boldsymbol{\delta}_3' - \frac{\pi}{2}\right)}\right] - \mu,$$

$$h_C(\mathbf{k}) = t'\left[e^{i\left(\mathbf{k}\cdot\boldsymbol{\delta}_1''+\frac{\pi}{2}\right)} + e^{i\left(\mathbf{k}\cdot\boldsymbol{\delta}_2''+\frac{\pi}{2}\right)} + e^{i\left(\mathbf{k}\cdot\boldsymbol{\delta}_3''-\frac{\pi}{2}\right)}\right] - \mu, \qquad (2)$$

where $\mu$ is the chemical potential of the fermion number. The off-diagonal SOC terms are

$$v_{AB}(\mathbf{k}) = i\lambda_R \sin(\mathbf{k}\cdot\mathbf{b}_1),\ v_{BC}(\mathbf{k}) = i\lambda_R \sin(\mathbf{k}\cdot\mathbf{b}_2),\ v_{CA}(\mathbf{k}) = i\lambda_R \sin(\mathbf{k}\cdot\mathbf{b}_3). \qquad (3)$$

We have $\mathbf{b}_1 = a(1,0)$, $\mathbf{b}_2 = a\left(-\frac{1}{2},\frac{\sqrt{3}}{2}\right)$, and $\mathbf{b}_3 = a\left(-\frac{1}{2},-\frac{\sqrt{3}}{2}\right)$. These correspond to three equivalent directions wrapping around the hexagonal center with uniform chirality. For symmetric Rashba hopping in KV$_3$Sb$_5$'s single triangle or minimal model, three $\mathbf{b}$ vectors for NNN paths suffice when spin-flip terms are sublattice-specific. However, six $\mathbf{b}$ vectors are essential for full tight-binding simulations of Kagome lattices with RSOC due to the multiple NNN paths connecting sublattice pairs. These vectors are given by

$$\mathbf{b}_1 = \mathbf{a}_1 + \mathbf{a}_2 = a\left(\tfrac{3}{2},\tfrac{\sqrt{3}}{2}\right),\quad \mathbf{b}_2 = -\mathbf{a}_1 + \mathbf{a}_2 = a\left(-\tfrac{1}{2},\tfrac{\sqrt{3}}{2}\right),\ \mathbf{b}_3 = -\mathbf{a}_1 - \mathbf{a}_2 = a\left(-\tfrac{3}{2},-\tfrac{\sqrt{3}}{2}\right),$$
$$\mathbf{b}_4 = \mathbf{a}_1 - \mathbf{a}_2 = a\left(\tfrac{1}{2},-\tfrac{\sqrt{3}}{2}\right),\ \mathbf{b}_5 = -\mathbf{a}_2 = a\left(-\tfrac{1}{2},-\tfrac{\sqrt{3}}{2}\right),\ \text{and}\ \mathbf{b}_6 = \mathbf{a}_2 = a\left(\tfrac{1}{2},-\tfrac{\sqrt{3}}{2}\right). \qquad (4)$$

The Kagome lattice's structure, with three sublattices and dual NNN paths between each pair of differing chirality, necessitates six unique $\mathbf{b}$ vectors to accurately model RSOC's effects on angular dependence and symmetry breaking.

## 3. Method

Our model of KV$_3$Sb$_5$ incorporates several essential physical mechanisms: NN hopping, complex NNN hopping with a phase factor $\phi = \pi/3$, RSOC, and a CDW amplitude. The resultant band structure, which emerges from three sublattices (A, B, C) combined with spin degrees of freedom (↑, ↓), comprises six distinct bands as in Fig.2. We need to lean on numerical analysis to obtain the eigenvalues of Eq.(1). We use the 'Matlab' package for this purpose. The numerical values of the eigenenergy and the corresponding eigenvectors are obtained by the command [V,D] = eig($H(\mathbf{k})$). The command returns diagonal matrix D of eigenvalues and matrix V whose columns are the corresponding right eigenvectors, so that $H(\mathbf{k})$ *V = V*D. For each k-point, in the chosen $k$-path, this process is repeated. The plots of the energy eigenvalues $E_j(k)$ ($j = 1,2,\ldots,6$) as ,obtained in this manner, are shown in Figure 2. The parameter values used are $t = 1, t' = 0.44$ (in Fig. a) $and\ t' = 0.86$ (in Fig. b), $\lambda_R = 0.14$ (in Fig. a) $and\ \lambda_R = 0.80$ (in Fig. b), $\Delta_{CDW} = 0.14$, and $J = 0.71$ (in Fig. a), and $J = 2.3$ (in Fig.b). We have plotted the band dispersion along high-symmetry lines in the 2D Brillouin zone ($\Gamma \to K \to M \to \Gamma$). The computed band structure reveals the characteristic flat band associated with kagome systems. These flat regions are attributed to destructive interference patterns arising from the lattice geometry and are known to promote electronic instabilities. Moreover, spin splitting induced by RSOC is readily apparent, emphasizing the role of spin–orbit interactions in lifting degeneracies. Importantly, in Fig.2b, the bands $E_3$ and $E_4$ ($E_5$ and $E_6$) display band-inversion near M-point (between M and K points). The introduction of a CDW leads to spectral gaps, indicating a reconstruction of the Fermi surface and breaking of translational symmetry. Within this structure, a saddle point, near the M- point, emerges that act as sources of enhanced Berry curvature. In fact, a saddle point in a band structure refers to a special point in the energy dispersion of electrons in a crystal leading to a van Hove singularity, manifesting as a distinct peak indicative of enhanced density of states due to band curvature. The resulting band structure also reveals asymmetric Dirac bands near the K-point, indicating broken inversion and time-reversal symmetry.

The quantum anomalous Hall conductivity (QAHC) may be calculated by integrating the Berry curvature on a k-mesh-grid of the Brillouin zone. The expression of QAHC is $\sigma_{xy} = -(\frac{e^2}{h}) \sum_n \int_{BZ} \frac{d^3k}{(2\pi)^3} f(E_n(k) - \mu) \Omega_n^z(k)$, where $\mu$ is the chemical potential of the fermion number, $n$ is the occupied band index, $f(E_n(k) - \mu)$ is the Fermi-Dirac distribution and $\Omega_n^z(k)$ is the z-component of the Berry curvature for the $n$th band. The latter could be calculated using the Kubo formula [19]

$$\Omega_n^z(k) = -2\hbar^2 \left[ Im \sum_{m \neq n} (E_n(k) - E_m(k))^{-2} \langle n, k | \widehat{v_x} | m, k \rangle \langle m, k | \widehat{v_y} | n, k \rangle \right]. \quad (5)$$

Here $\boldsymbol{k}$ is the Bloch wave vector, $E_n(k)$ is the band energy, $|n, k\rangle$ are the Bloch functions of a single band. The operator $\hat{v}_j$ represents the velocity in the *j* direction. We recall that for a system in a periodic potential and its Bloch states as the eigenstates, in view of the Heisenberg equation of motion $i\hbar \frac{d\hat{x}}{dt} = [\hat{x}, \hat{H}]$, the identity

$$\langle m, \mathbf{k'}|v_\alpha|n, \mathbf{k}\rangle = \left(\frac{1}{\hbar}\right) (E_n(\mathbf{k'}) - E_m(\mathbf{k})) \left\langle m, \mathbf{k'}\left|\frac{\partial}{\partial k_\alpha}\right|n, \mathbf{k}\right\rangle \qquad (6)$$

is satisfied. Upon using this identity, we obtain Hall conductivity in the zero temperature limit as $\sigma_{xy} = C\left(\frac{e^2}{\hbar}\right)$ where $C = \sum_n C_n,\ C_n = \int \int_{BZ} \Omega_{xy}(k) \frac{d^2k}{(2\pi)^2}$. The z-component of the Berry-curvature(BC) is

$$\Omega_{xy}(k) = \left(\frac{\partial A_{n,y}}{\partial k_x} - \frac{\partial A_{n,x}}{\partial k_y}\right) = -2\ Im \left\langle \frac{\partial \psi_{n,\mathbf{k}}}{\partial k_x}\bigg|\frac{\partial \psi_{n,\mathbf{k}}}{\partial k_y}\right\rangle \qquad (7)$$

where $\psi_{n,\mathbf{k}} = |n, k\rangle$. The vector potential $\mathbf{A}_n(\mathbf{k})$ is the Berry connection and $\nabla_\mathbf{k} \times \mathbf{A}_n(\mathbf{k}) = \Omega_n(\mathbf{k})$ is derived from it. While the derivative calculations involved in the Kubo formula are conceptually straightforward, they are computationally intensive. To investigate whether the Chern number in our system is quantized, we adopted the Fukui-Hatsugai-Suzuki (FHS) method **[14–17]**, which assumes a discretized Brillouin zone. This technique is grounded in lattice gauge theory, replacing continuous derivatives with discrete link variables. Crucially, the FHS method preserves gauge invariance under lattice boundary conditions. The Chern number is extracted from the Wilson loop, defined as the product of link variables around a plaquette in momentum space. This formulation remains invariant under local phase transformations of the Bloch wavefunctions, ensuring robust topological characterization.

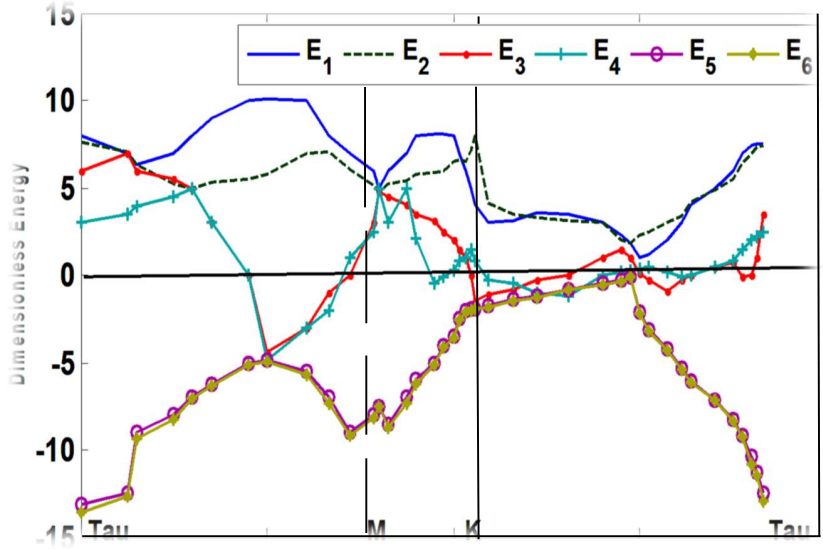

(a)

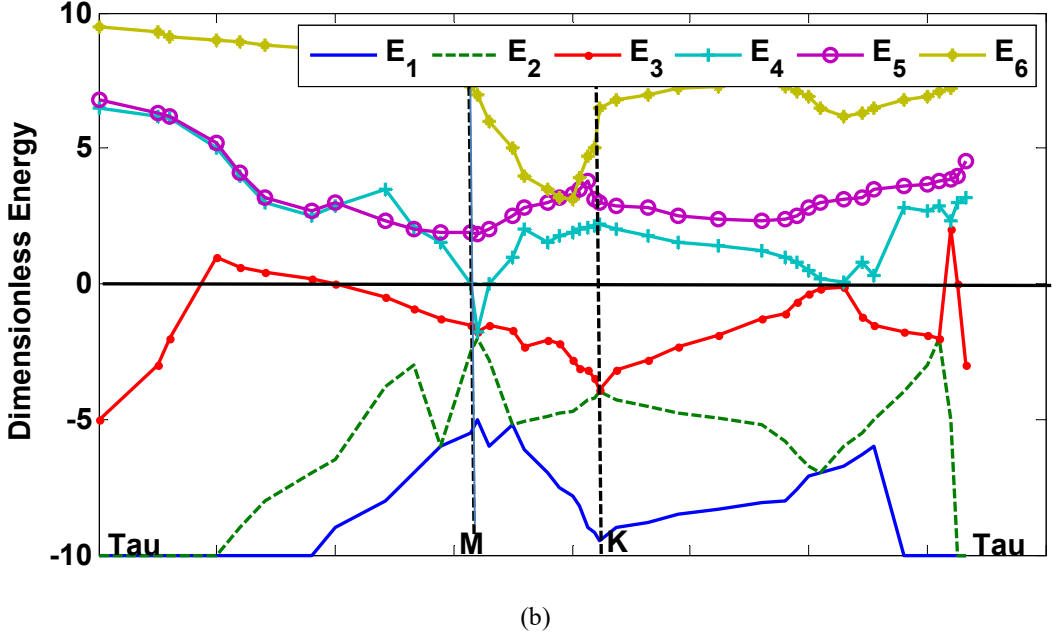

(b)

**Figure 2. (a) (b)** The (six) band structure for the Kagome lattice model of KV$_3$Sb$_5$ using NN hopping, NNN hopping (complex with phase $\phi = \pi/3$ in *a*, but $\phi = \phi(\mathbf{k}) = (\sin(k_x) - \sin(k_y))$ in *b* ), RSOC, and CDW amplitude. The six bands arise from the combination of three sublattices (A, B, C) and two spin states (↑, ↓). The high-symmetry path followed is $\Gamma \to M \to K \to \Gamma$. The parameter values used are $t = 1, t' = 0.44\ (in\ Fig.a)\ and\ t' = 0.86\ (in\ Fig.b), \lambda_R = 0.14 (in\ Fig.a) and\ \lambda_R = 0.80\ (in\ Fig.b), \Delta_{CDW}= 0.14,\ and\ J = 0.71(\ in\ Fig.a), and\ J = 2.3 (in\ \ Fig.b)$. The graphical representation exhibits the spin splitting due to SOC. The spectral gaps are induced by CDW. The figure exhibits a band structure with saddle points, leading to Berry curvature hotspots—key ingredients for non-zero Chern numbers. There is near-flat band feature close to the Fermi level.

A quantitative topological analysis was conducted by evaluating the Berry curvature of individual bands on a coarse $50 \times 50$ momentum-space grid using the parameter values as $t = 1$, $t' = 0.44$, $\lambda_R = 0.14$, $J = 0.71$, and $\Delta_{CDW} = 0.14$ in Fig. 3. We obtain the following values of the Chern number $C_1 = +0.00$, $C_2 = -0.02$, $C_3 = +0.03$, $C_4 = +0.09$, $C_5 = +0.01$, and $C_6 = -0.04$ corresponding to $E_1$, $E_2$, $E_3$, $E_4$, $E_5$, and $E_6$, respectively, in Fig.2a. The NNN hopping is complex with phase $\phi = \pi/3$. The results hint at possible topological transitions and potential

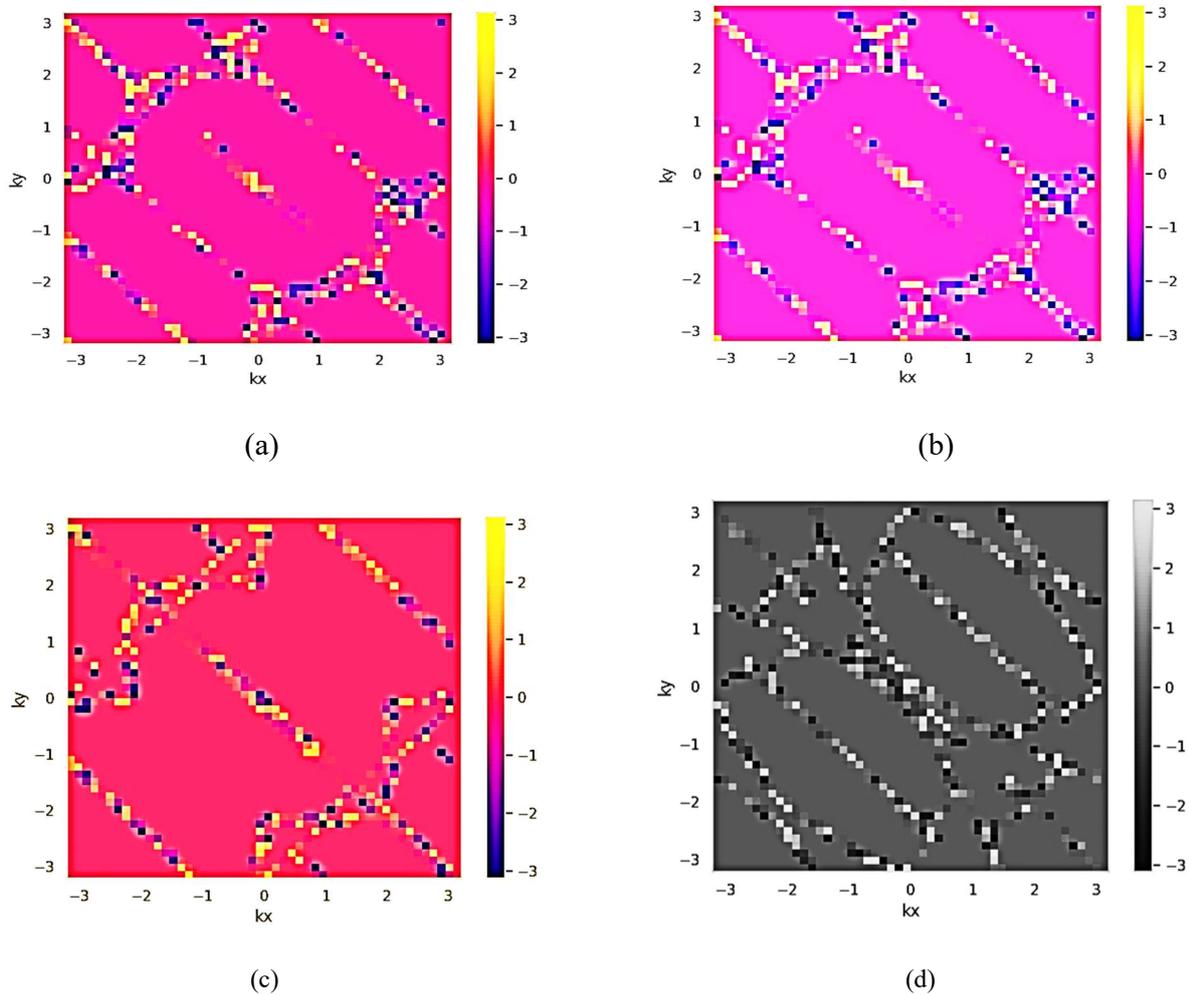

(a)

(b)

(c)

(d)

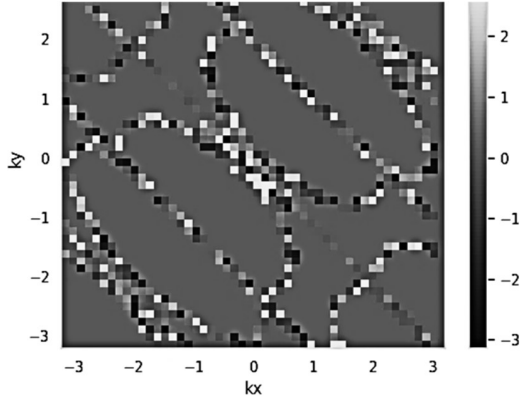
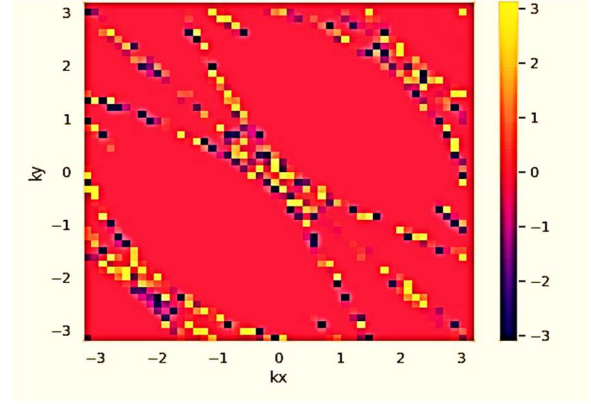

(e)  (f)

**Fig.3.** The 2D plots of the Berry curvature, in Figs.3a, 3b, 3c, 3d, 3e, and 3f , as a function of $k_x$ and $k_y$ corresponding to the energies $E_1$ , $E_2$, $E_3$, $E_4$, $E_5$, and $E_6$ in Fig.2a, respectively. The parameter values used are $t = 1$ , $t' = 0.44$, $\lambda_R = 0.14$, $J = 0.71$ , and $\Delta_{CDW}= 0.14$. The NNN hopping is complex with phase $\phi = \pi/3$. We obtain the following values of the Chern number $C_1 = +0.00$, $C_2 = -0.02$, $C_3 = +0.03$, $C_4 = +0.09$, $C_5 = +0.01$, and $C_6 = -0.04$ corresponding to $E_1$, $E_2$, $E_3$, $E_4$, $E_5$, and $E_6$, respectively, in Fig.2a.

Chern band formation with tweaking of parameters. In other words, these create a fertile ground for further exploration of topological transitions, though the current scenario has not yet crossed the threshold for integer Chern numbers. In Fig.4, we have the 2D plots of the Berry curvature, in Figs. a, b, c, d, e, and f , as a function of $k_x$ and $k_y$ corresponding to the energy bands $E_1$ , $E_2$, $E_3$, $E_4$, $E_5$, and $E_6$ in Fig.2b, respectively.

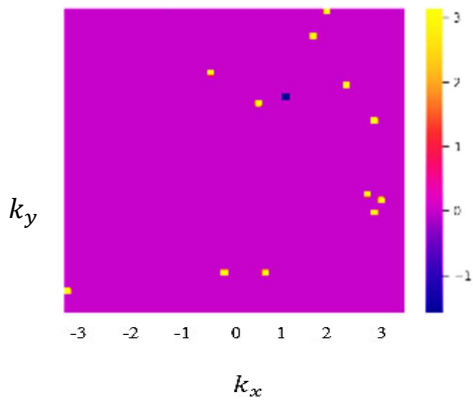
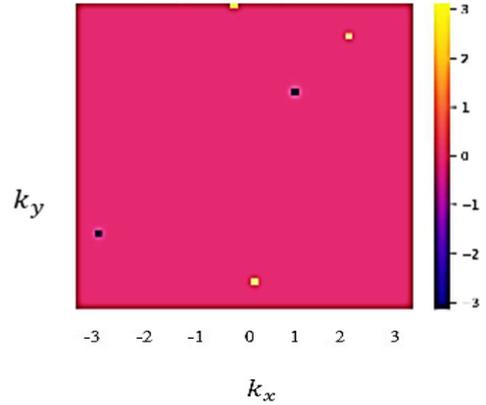

(a)  (b)

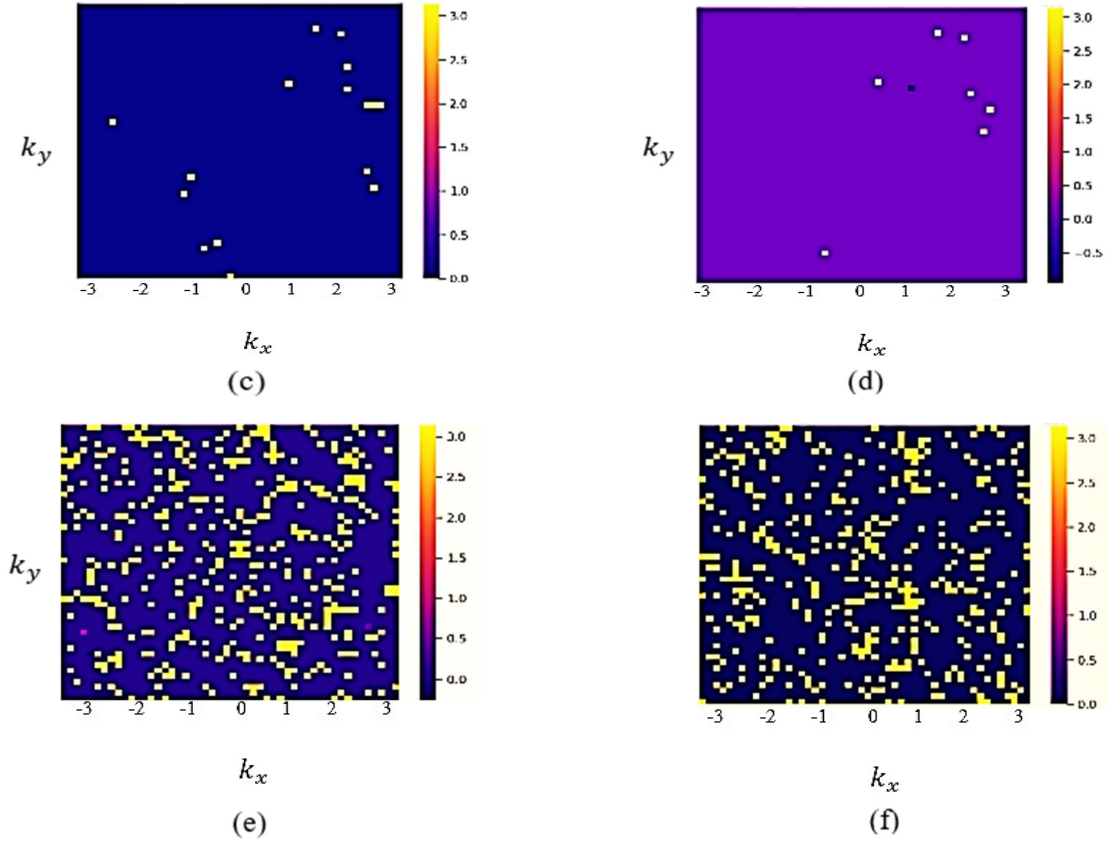

**Fig.4.** The 2D plots of the Berry curvature, in Figs. 4a, 4b, 4c, 4d, 4e, and 4f, as a function of $k_x$ and $k_y$ corresponding to the energy bands $E_1$, $E_2$, $E_3$, $E_4$, $E_5$, and $E_6$ in Fig.2b, respectively. The Chern number for the Berry curvature (Fig. 4a) of the band $E_3$ (Fig. 2b) is $C_1 \approx +1$, and that (Fig. 4b) for the band $E_4$ (Fig. 2b) is $C_2 \approx -1$. The rest of the bands (Berry curvature plots of which are shown in Fig. 4c, 4d, 4e, and 4f) have $C \approx 0$. The numerical value of the parameters used are $t = 1$, $t' = 0.86, \lambda_R = 0.8, J = 2.3$, and $\Delta_{CDW} = 0.14$. The phase ϕ of the complex NNN hopping ($t'$) is assumed to be momentum-dependent: ϕ(k) = (sin(k$_x$) – sin(k$_y$)). This introduces momentum-space winding, mimicking an orbital magnetic flux.

The Chern number for the Berry curvature (Fig. 4a) of the band $E_3$ (Fig. 2b) is $C_1 \approx +1$, and that (Fig. 4b) for the band $E_4$ (Fig. 2b) is $C_2 \approx -1$. The rest of the bands (Berry curvature plots of which are shown in Fig. c, d, e, and f) have $C \approx 0$. The numerical value of the parameters used are $t = 1$, $t' = 0.86, \lambda_R = 0.8, J = 2.3$, and $\Delta_{CDW} = 0.14$. The phase ϕ of the complex NNN hopping ($t'$) is assumed to be momentum-dependent: ϕ(k) = (sin(k$_x$) – sin(k$_y$)). This introduces momentum-space winding, mimicking an orbital magnetic flux. With Rashba spin-orbit coupling cranked up to $\lambda_R = 0.8$ and exchange field strength boosted to $J = 2.3$, our KV$_3$Sb$_5$ system is entering a regime of strong spin splitting and magnetic asymmetry. Combined with the momentum-dependent ϕ, this setup flattens bands and intensifies Berry curvature hotspots. This is huge: Bands 3 and 4 now carry opposite Chern numbers, indicating the emergence of chiral edge states and the possibility of quantized anomalous Hall effect (QAHE). The rest remain trivial, but the system as a whole is no longer topologically inert. We, thus, have a system with chiral edge modes and, more importantly, a platform for topological device engineering.

## 4. Future perspective and concluding remarks

Given that our analysis reveals two occupied bands with opposite Chern numbers, a natural question arises: how can such a system support a quantized anomalous Hall effect (QAHE)? To address this, we propose a van der Waals heterostructure (VDWH) composed of a $KV_3Sb_5$ layer and a monolayer transition metal dichalcogenide (TMD), such as $MoS_2$ or $WSe_2$. The $KV_3Sb_5$ layer contributes six bands ( 6 by 6 Hamiltonian: $[H_{KV_3Sb_5}(\boldsymbol{k})]$ characterized by strong Berry curvature and potential topological features, while the TMD layer contributes six bands [6 by 6 Hamiltonian: $T(\boldsymbol{k})$] derived primarily from the d-orbitals of the transition metal, exhibiting strong spin-orbit coupling (SOC) and valley-dependent physics. In monolayer TMDs, the valence bands predominantly originate from the $d_{x^2-y^2}$ and $d_{xy}$ orbitals, while the conduction bands are mainly derived from the $d_{z^2}$ orbital. Strong SOC, particularly in tungsten-based compounds, leads to significant spin_ splitting. At each valley (K and K′), this results in two spin-split valence bands and two spin-split conduction bands. Additional bands may arise from higher or lower energy d-orbitals or through substrate-induced hybridization. Thus, it is reasonable to identify six distinct bands in the TMD layer that exhibit strong SOC and valley-selective behavior. The feasibility of such a band structure within a vHs stems from favorable band alignment and interlayer coupling. The VDW interface preserves the intrinsic electronic properties of each constituent layer while allowing for proximity-induced effects. Notably, the SOC from the TMD layer can influence the electronic states of $KV_3Sb_5$, and vice versa. Furthermore, the breaking of inversion and/or time-reversal symmetry at the interface can enhance Berry curvature, potentially enabling the emergence of QAHE when magnetic ordering is introduced. This heterostructure provides a 12-band basis, leading to a momentum-space Hamiltonian defined on a 12×12 matrix:

$$[H_{KV_3Sb_5}(\boldsymbol{k}) \quad T(\boldsymbol{k}) \\ T^\dagger(\boldsymbol{k}) \quad H_{KV_3Sb_5}(\boldsymbol{k})] \tag{8}$$

This framework sets the stage for exploring topological phases and spin-valley coupled transport phenomena in engineered quantum materials.

As outlined in Sect. 2, the coexistence of chiral CDW order, $Z_3$ nematic symmetry breaking, and field-tunable chirality makes $KV_3Sb_5$ a compelling platform for exploring quantum phenomena such as topological phases, unconventional superconductivity, and electronic symmetry breaking. The ability to reversibly control collective electronic motion via magnetic flux highlights its promise for advancing quantum materials research and enabling future technologies based on topological manipulation and quantum coherence. Our upcoming work focuses on probing unconventional superconductivity in $KV_3Sb_5$ and investigating a van der Waals heterostructure (VDWH) comprising a TMD monolayer and $KV_3Sb_5$. Preliminary findings suggest that this hybrid system may support the quantum anomalous Hall effect (QAHE), offering exciting prospects for topological electronics.

We provided in this paper an explicit example of the Bloch Hamiltonian matrix in momentum space for a Kagome lattice of $KV_3Sb_5$ with one orbital, one layer, three sublattices (A, B, C), and two spin states (↑, ↓). It helps visualize the structure without being overly large. This setup gives the six-band(s) structure. We have calculated the Chern numbers for the bands. We find the possibility of quantized anomalous Hall effect (QAHE) albeit with two bands carrying opposite Chern numbers. We have indicated above that the proposed hybrid system may support QAHE. These are the highlights. The experimental realization of QAHE is yet to be achieved due to various theoretical and practical obstacles. Specifically, the following challenges exist: (a) $KV_3Sb_5$ does not exhibit intrinsic magnetism in the conventional sense. TRS breaking may occur due to orbital or CDW-induced effects, rather than magnetic spins. (b) QAHE requires the emergence of a bulk gap at Dirac points due to magnetic interactions, which has not been clearly observed in $KV_3Sb_5$. (c) QAHE necessitates clean systems, band isolation, and often fine-tuned chemical potentials. Real-world $KV_3Sb_5$ samples may be affected by excessive disorder or imperfect band alignment.

**References:**

**1.** E. Uykur, B.R. Ortiz, *et al.*, *npj Quantum Mater.* **7**, 16 (2022). https://doi.org/10.1038/s41535-021-00420-8

**2.** Z. Guguchia, C. Mielke, D. Das, *et al.*. *Nat Commun* **14**, 153 (2023). https://doi.org/10.1038/s41467-022-35718-z

**3.** M Michael Denner, Ronny Thomale, *et al.*. Phys. Rev. Lett. 127, 217601(2020). Erratum: Phys. Rev. Lett. 128, 099901(2022). https://doi.org/10.1103/PhysRevLett.127.217601

**4.** Yu-Xiao Jiang, *et al.*, arXiv:2012.15709 [cond-mat. supr-con](2020).

**5.** M Zahid Hasan *et al.*, Nat Mater. Oct;20(10) :1353-1357 (2021). https://doi.org/ 10.1038/ s41563-021-01034-y

**6.** J. Zhao, S. A. Yang *et al.*, arXiv:2103.15078 v2 [cond-mat. str-el.](2021).

**7.** H. Luo, Q. Gao, H. Liu *et al.*, *Nat Commun* **13**, 273 (2022). https://doi.org/10.1038/s41467-021-27946-6

**8.** S. Cheng, Z. Ren *et al.*, arXiv:2302.12227(2023). DOI:10.48550/arXiv.2302.12227

**9.** M.L. Kiesel, C. Platt, and R. Thomale, Phys. Rev. Lett. 110, 126405 (2013).

**10.** W.-S. Wang, Z.-Z. Li, *et al.*, Phys. Rev. B 87, 115135 (2013).

**11.** X. Zhu, Y. Cao, *et al.*, Proc. Natl Acad. Sci. USA112, 2367 (2015).

**12.** Y. Cai, Y. Wang, Z. Hao *et al.* *Commun Mater* **5**, 31 (2024). https://doi.org/10.1038/s43246-024-00461-z

**13.** I. Makhfudz, M. Cherkasskii, *et al.*, Phys. Rev. B 110, 235130 (2024).

**14.** T. Fukui, Y. Hatsugai, and H. Suzuki, J. Phys. Soc. Jpn. 74, 1674 (2005).

**15.** U.P. Tyagi, and Partha Goswami, Acta Physica Polonica A, Vol. 148 No. 1 (2025).

**16.** M. Nakamura and S. Masuda, Phys. Rev. B 110, 075144 (2024).

**17.** T. Shiina, F. Hamano, and T. Fukui, Phys. Rev. B 111, 245135(2025).

**18.** Partha Goswami and U.P. Tyagi, Acta Physica Polonica A, Vol. 144 No. 3 (2023).

**19.** M. V. Berry, Proc. Roy. Soc. Lond. A 392, 45 (1984).